# Application of non-uniform Fourier transform to non-uniform sampling Fourier transform spectrometers

Authors:

Muqian Wen, John Houlihan

## Abstract

Resampling by interpolation is the traditional method to process interferograms from non-uniformly sampled Fourier transform spectrometers. The non-uniform fast Fourier transform (NUFFT) is an alternative approach that has been mostly overlooked. With the aid of experiments on a high-resolution interferometer with a variety of optical sources, these two methods are compared. It is found that the NUFFT is comparable to interpolation in spectral profile shape and spectral noise levels and is better in spectral amplitude and computer performance. A significant advantage is also found in the case of under-sampling and noise performance by NUFFT due to the unique non-periodic nature of non-uniform sampling. In addition, a novel implementation of NUFFT is presented and analysed.

Keywords: non-uniform Fourier transform, Fourier transform spectrometer, Michelson interferometer, non-uniform sampling, NUFFT

## 1. Introduction

The Michelson interferometer and its many variants are the foundation of many optical and non-optical spectrometry designs. One particular type is the Fourier transform spectrometer [1]. It uses a moving mirror in one of the arms of Michelson interferometer to generate interference signals and uses a single pixel photodetector to record the interferogram. It is one of the most common types of optical spectrometers and has seen applications up to the ultraviolet range at 40nm wavelength [2]. Although its principle is relatively simple, it has many advantages. It can allow relatively large percentage of light to reach the photodetector, records all frequency component simultaneously, has intrinsic wavelength calibration and is able to achieve high wavelength resolutions with relative ease. It is especially advantageous in infrared region where it is one of the most popular techniques.

One of the challenges in Fourier transform spectroscopy is the difficulty in obtaining evenly spaced samples in optical path difference space due to the difficulty to precisely maintain constant mirror moving speed. One of the simplest solutions is to sample non-uniformly instead and then use interpolation to resample the data uniformly such as the example here [3]. This method has the advantage of minimising hardware requirements and can be readily realized with off-the-shelf optical components and thus is suitable for many situations. However, in such cases of non-uniform sampling Fourier transform interferometers there also exists another overlooked method to process the data into spectrum which is to use the non-uniform fast Fourier transform (NUFFT).

The NUFFT is a concept that has been discussed in many areas in recent decades [4]. It has many applications in a diverse range of fields ranging from magnetic resonance imaging(MRI) [5], ultrasound imaging [6] and radar imaging [7], to numerical solutions of differential and integral equations [8], filter design [99], electron microscope image alignment [10], to mention a few. It has been used and proven superior in various interferometric applications such as spectral domain optical coherence tomography (SD-OCT)[11], holography diffraction calculation [12], wavelength-tuning interferometry (WTI)[13] and static single-mirror Fourier transform spectrometer (sSMFTS)[14], etc.

However, while the NUFFT has been seen extensively in more complicated interferometric systems, to our best knowledge its application in optical spectrometry systems has been very rare and has not been experimentally reported in the simplest type of Michelson interferometer-based spectrometers, the single dimensional moving mirror optical Fourier transform spectrometer. There has been one theoretical study on the comparison of NUFFT with the interpolation method [15] which shows numerically that NUFFT method is superior to the interpolation method in simulation in spectrum emission/absorption line amplitude, width, position etc. However, this behaviour has yet to be studied experimentally. Studying the application of NUFFT to this simplest spectrometry case of non-uniform sampling may not only help improve the technology of Fourier transform spectroscopy but can also have implications for other applications that employs interferometers. Therefore, this paper will report an experimental study of the application of the NUFFT to the Michelson interferometer and outline the various related advantages and disadvantages.

The experimental study involved the building of a high resolution nonuniform sampling Fourier transform spectrometer to systematically study the performance and differences of NUFFT method over resampling by interpolation FFT method. The visible and IR ranges are investigated by using the interferograms of a 532nm solid-state multimode laser, a 960nm broad spectrum SLED as well as a 632.8nm Helium Neon laser reference source. This interferometer has a scanning length of about 5 meters thus allowing the NUFFT performance at very high wavelength resolutions to be evaluated.

## 2. Theory

**2.1 Non-uniform Fourier transform**:

Non-uniform discrete Fourier transform theory (NUDFT) is the non-uniform variant of the regular discrete Fourier transform (DFT). It is defined by the following equation:

$$F(v_k) = \sum_{n=1}^{N} f(x_n)e^{-i2\pi v_k x_n} \qquad k = 1,2 \ldots M$$

where $v_k$ represents frequency at the k-th spectral point, $x_n$ represents the position of the *n*-th sampling point, $f(x_n)$ represents the signal strength at the sampling position $x_n$, and $F(v_k)$ represents the spectral strength at the frequency $v_k$. In fact, the discrete Fourier transform can be considered a special case of non-uniform Fourier transform when all positions and frequencies are distributed uniformly. When all points are distributed uniformly the equation of nonuniform Fourier transform becomes the same as the equation of discrete Fourier transform.

Direct computation of the discrete Fourier transform is considered prohibitively costly because the computation time scales with the square of number of sample points and instead the fast Fourier transform (FFT) is usually employed. The NUFFT and FFT are the fast algorithms of NUDFT and DFT respectively. Theoretically, the FFT speed is proportional to $O(N \log N)$ where N is the number of sample or spectral points and NUFFT speed is also similarly proportional to $O(N \log N + M)$ where N is the number of spectral points and M is the number of sample points [16,17].

In this paper the NUFFT calculation will be performed using the python FINUFFT package by the Flatiron Institute [16,17].

**2.2 Resampling by Interpolation and FFT**:

When an experiment yields a non-uniform sampled interferogram, the conventional approach is to uniformly resample the data by using interpolation and then use the regular discrete Fourier transform method (or FFT) to process the data. The equation of a regular discrete Fourier transform is given below which is actually a special case of nonuniform Fourier transform:

$$F(v_k) = \sum_{n=1}^{N} f(x_n) e^{-i2\pi \frac{kn}{N}} \quad k = 1,2 \ldots N$$

For interpolation, it has been reported that cubic spline interpolation gives better results than linear interpolation in this application [15]. We also find that it gives better results, so we follow this approach throughout. In this paper the interpolation and FFT calculations are based on the SciPy [18] python package using its interpolation module and FFT module.

It should be noted that both NUFFT and interpolation / FFT approaches require prior knowledge of the positions of the sampling points. This paper would use a monochromatic reference helium neon laser to determine the positions of sampling points whose principle will be described below. We use this method because of its simplicity and because it allows us to use a single photodetector only. Similar methods based on this principle of using a reference monochromatic source have been reported before [19,20,21]. It should be noted that this is by no means the only way to determine sample positions as attested by the two examples here which use totally different methods that do not require reference source at all [3,14].

For an ideal monochromatic light, the interference signal will simply be a cosine function with its phase being proportional to the optical path difference. Thus, the relative optical path difference can be easily determined by simply doing inverse cosine calculation of the reference signal. In practice however the situation is more complicated because the amplitude of the signal will vary due to many reasons such as the divergence of the light beam as it propagates, slight misalignments of the movable mirror during its movement, intrinsic variations in output intensity of the laser in time, and ambient light pollutions etc. To address this issue, a method based on instantaneous phase of the analytic signal is chosen for this experiment. The analytic signal is a complex valued signal containing only the positive frequency part of the spectrum of the real signal. The spectrum of a real value is always symmetric with equal negative and positive parts. Thus, the analytic signal will always contain the same information as the real signal, and it can be obtained by inverse Fourier transformation of the positive part of spectrum. The real time domain interference signal of a monochromatic light source with varying mirror moving speed and fluctuating signal amplitude can be written as:

$$f(t) = A(t) \cos(\omega x(t)) = A(t)(e^{i\omega x(t)} + e^{-i\omega x(t)})$$

where $f(t)$ represents the signal, $A(t)$ represents signal amplitude which is a real value function, $x(t)$ represents the optical path difference which is a function of time, $\omega$ represents the frequency of this monochromatic signal in optical path difference domain, and $t$ represents time. The analytic signal $z(t)$ in time domain of this real signal can then be obtained by removing the negative frequency component from the original signal $f(t)$ and in this case it can be assumed to be of the following form which has its instantaneous phase proportional to optical path difference:

$$z(t) = A(t)e^{i\omega x(t)}$$

As a result, the amplitude variations can now be easily separated from phase calculation. Hence, the relative optical path difference can be determined from instantaneous phase of the analytic signal if the wavelength of the reference source is also known. In addition, this method has the benefit of potentially allowing only a single photodetector to be used thus further simplifying hardware design.

# 3. Experiment

**3.1 Design:**

In this paper we design and build an unconventional variant of Michelson interferometer (Figure 1) in order to achieve a very long scanning length. The classical Michelson interferometer would have one stationary arm and another mobile arm with moving mirrors. However, in this experiment, we use two mobile arms moving in opposite directions to double the optical path difference. The interferometer is built on a 306mm translation stage and has an effective scanning length close to 5 meters while by comparison the highest resolution Fourier transform spectrometer we found from the literature the Bruker IFS 125HR has a scanning length of about 11 meters. In order to achieve this scanning length, four retroreflectors are employed to fold the optical paths at each arm by 8 times and we place the moving mirror of both arms onto the translation stage in back-to-back fashion to further increase the optical path difference by two times and thus resulting in 16 times increase in optical path differences in total. A reference 632.8nm helium neon laser is added to the test light source using a beam splitter to determine the position of the sampling points.

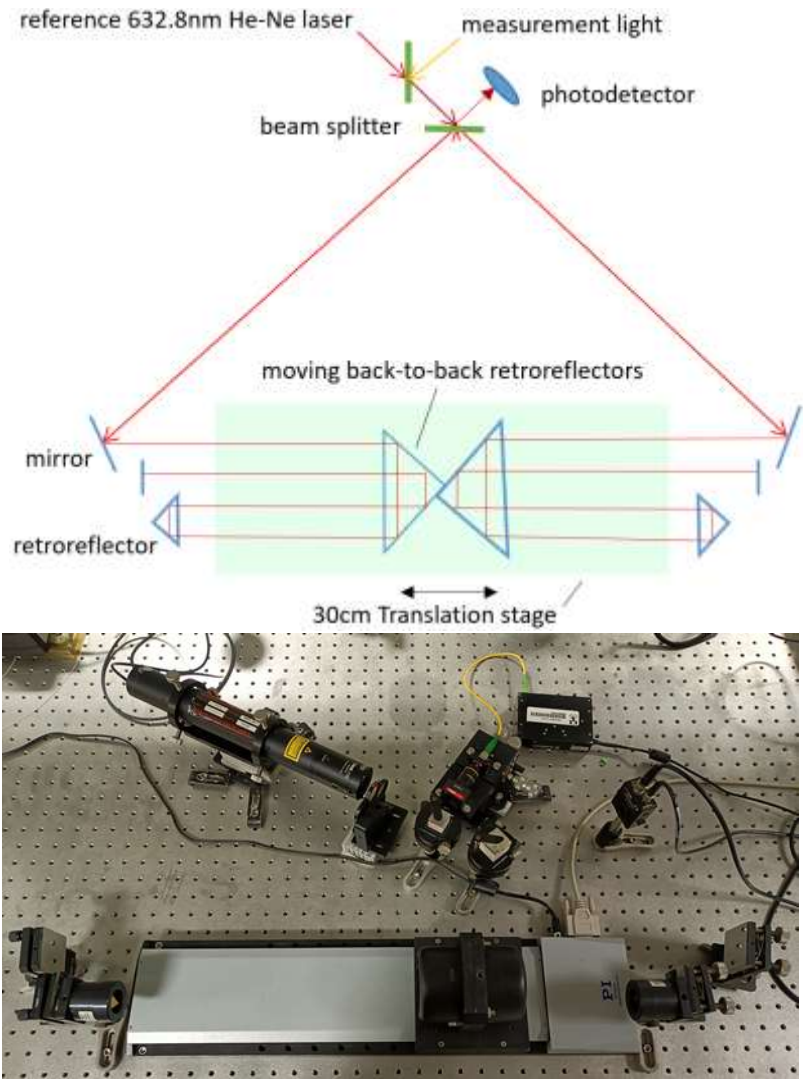

*Figure 1 Schematic (top) and photgraph (bottom) of the interferometer used in this experiment*

We use this interferometer the record the interferogram of a 532nm multimode solid-state laser and a 960nm broad spectrum SLED infrared light source as well as the 632.8nm Helium Neon laser reference source.

The various light sources used by this experiment are as follows: Thorlabs HRS015 frequency stabilized 632.8nm helium neon laser, Superlum SLD-MS-261 960nm SLED, CNI MGL-III-532-20mW 532nm solid state laser. The photodetector used here is Thorlabs PDA100A-EC 340nm-1100nm Switchable Gain Detector. The data acquisition device used is National Instrument NI-PCI-MIO-16E-4 with a maximum sampling rate of 1666666Hz. The translation stage used is a Physik Instrumente M-531.DDX translation stage with Mercury C-860.10 DC-motor controller.

An example of the interferogram of the 532nm laser and 960nm SLED is provided below (Figure 2) to demonstrate the signal characteristics of these test light sources. It can be seen as expected that the coherence length of the laser is very long while the coherence length of the SLED is very short. The interferograms are obtained with the reference helium neon laser turned off and the optical path difference is estimated from average mirror moving speed.

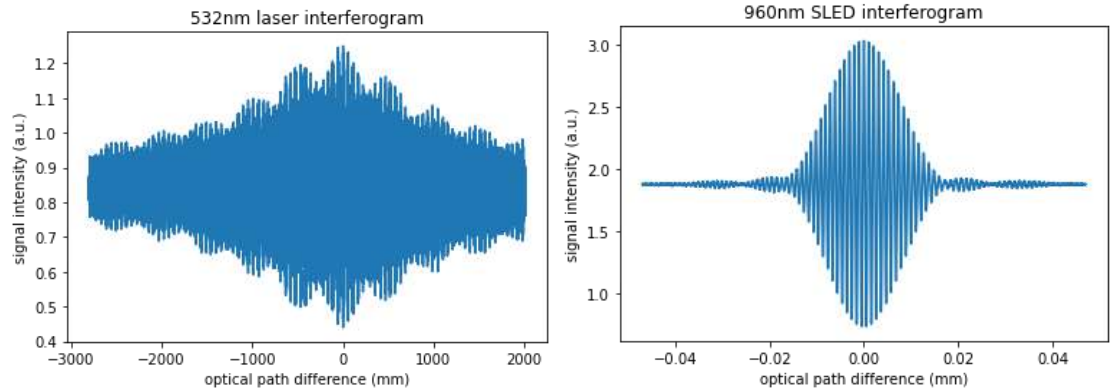

*Figure 2 An example of the interferogram of the 532nm laser (left) and the 960nm SLED (right). The optical path differences are estimated from average mirror moving speed.*

### 3.2 Characteristics:

Figure 3 shows an example of the raw interferogram of the helium neon laser and a zoom in segment of it as well as the positive part of the Fourier transform of this interferogram. The Fourier transform profile is shown to demonstrate the effect of mirror moving speed variations.

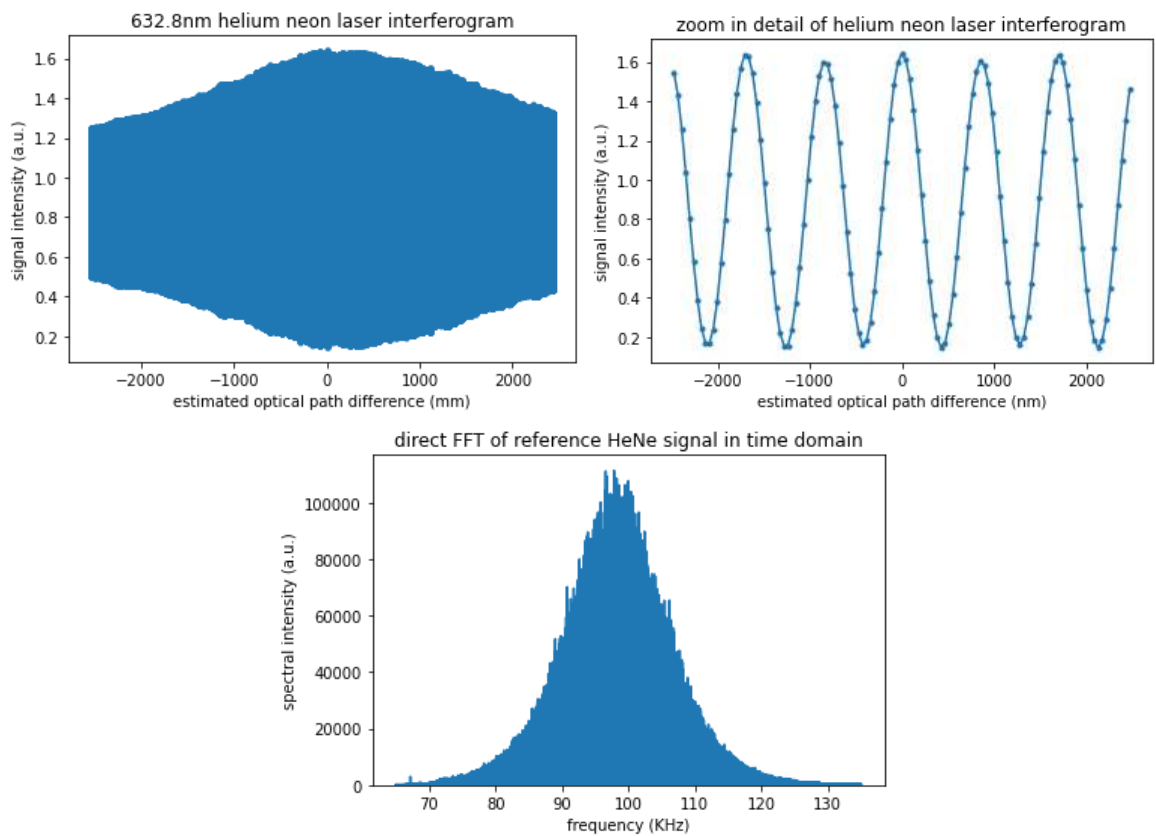

*Figure 3 An example of the raw interferogram (top left) of the helium neon laser, a zoom-in detailed segment of it (top right) and the positive part of the corresponding Fourier transform spectrum (bottom).*

Figure 4 shows the calculated position profile of the sample points. As outlined in the previous section, the position is directly proportional to the instantaneous phase of the real interferogram obtained from the complex analytic signal. To examine the variability in sampling, a sampling "speed" in optical path difference domain is obtained by differentiation of this sampling position profile. As can be seen from Figure 4 the sampling position shows a mainly linear behaviour, however the graph of sampling speed indicates large speed variations as much as 30%.

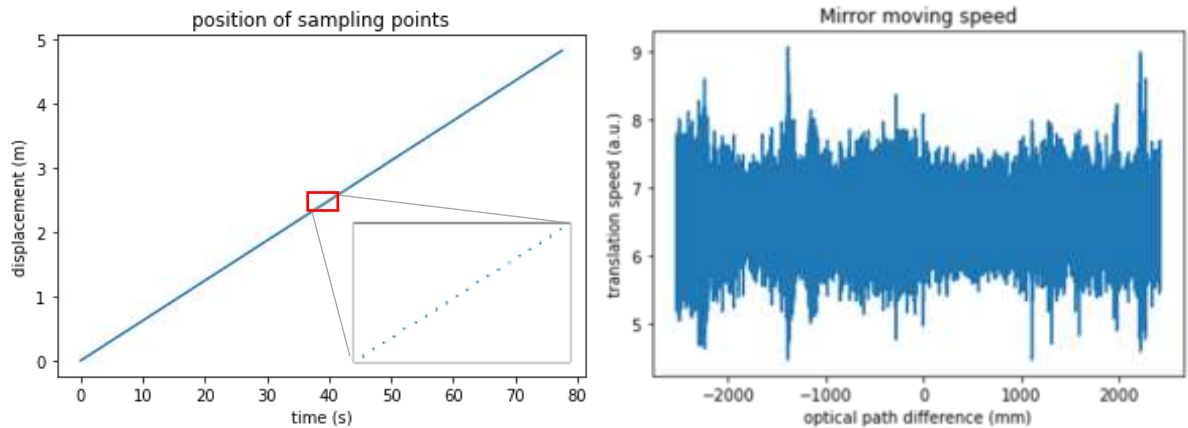

*Figure 4 An example of the sampling point position profile (left) and the calculated translation stage speed profile (right). The translation stage is set to move at about 4mm/s in this example and the sampling rate is 1500KHz.*

## 4. Results

### 4.1 Spectral profile comparison:

First, we compare the spectrum profile of the 960nm SLED and 512nm laser calculated from both methods. It should first be noted here that there is one technical difference between NUFFT and interpolation method regarding the handling of 0Hz or DC component in obtained interferogram. Raw unfiltered interference signals will always contain a large DC component due to the nature of interference. For the regular or interpolation based FFT method this component will not affect other parts of the spectrum and can simply be ignored. However, for the NUFFT this 0Hz component can in theory generate noise in the rest of spectrum and so removing this component first before NUFFT calculation can in some situations significantly improve the spectral result. We use an interferogram of the 960nm SLED (Figure 5) to demonstrate this phenomenon which demonstrated that the large impact on spectrum quality this effect can have. In general, this impact is less pronounced when the sampling rate is higher or when the spectral amplitude is relatively high such as in the case of a laser source.

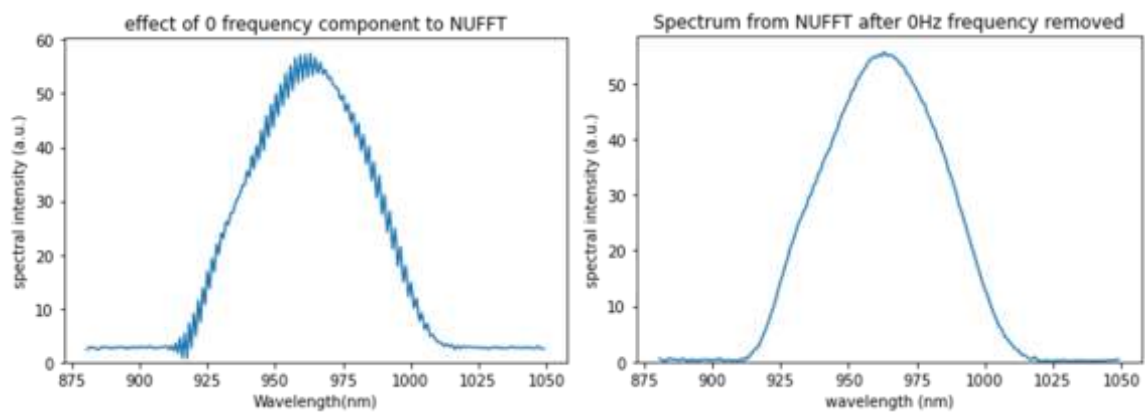

*Figure 5 Spectral noise in NUFFT spectrum due to presence of 0Hz component. The sampling interval is about 105nm.*

While this problem may seem trivial as the 0Hz component can be easily removed by simply subtracting the mean value of the interferogram, this issue is important in a practical sense as this source of error can be very tricky to identify without knowing it in advance. The reason for this behaviour is that the 0Hz component amplitude is much larger than the amplitude of the actual interference wave components.

### 4.1.1 General profile shape:

Next, we show examples of the calculated spectra of the 960nm SLED and 532nm laser sources (Figure 6 and Figure 7 respectively). In both SLED and laser cases, the two methods produce almost identical spectral profiles in terms of overall profile shape. In these examples, the NUFFT spectrum amplitude is about 5% higher than that of interpolation method for 960nm SLED and about 0.5% higher for the 532nm laser case. The dependence of these percentage on source type / spectral method will be further investigated in the later section 4.1.3.

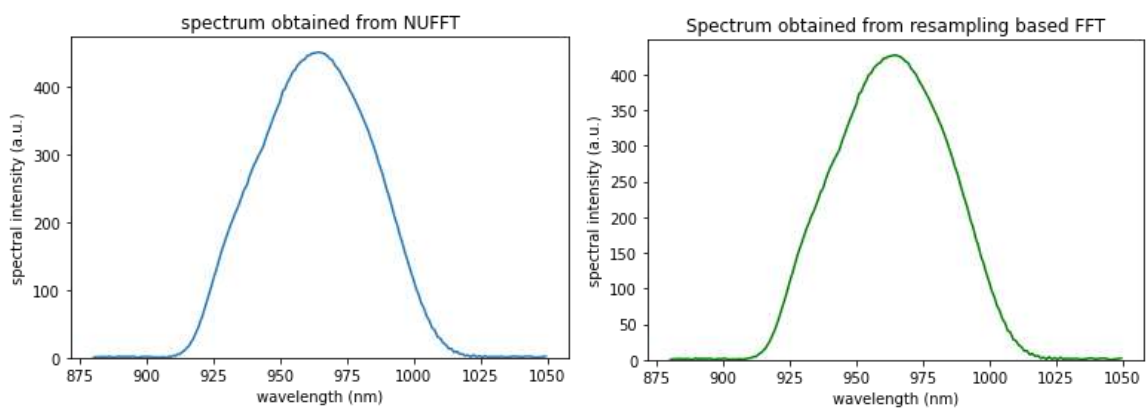

*Figure 6 Comparison of the NUFFT (left) and interpolation FFT method (right) obtained spectrum of the 960nm SLED source. The translation stage average moving speed is 2mm/s while sampling rate is 1500KHz.*

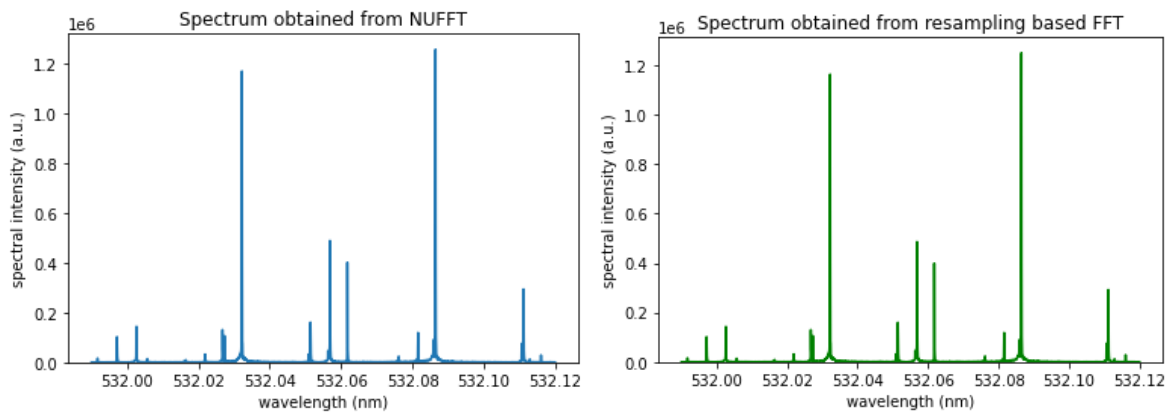

*Figure 7 Comparison of the NUFFT (left) and interpolation FFT method (right) obtained spectrum of the 532nm multimode laser. The translation stage average moving speed is 11mm/s while sampling rate is 1666666Hz.*

### 4.1.2 Spectral noise:

To compare the spectral noise levels for both conventional and NUFFT methods, Figure 8 contains the 532nm laser with reference He-Ne spectra on a log scale. Again, here it can be seen that their spectral noise profiles are very similar when the sampling rate is around 1 sample per 100nm. The mean to peak ratio of the log value of the spectrum is 0.162 for NUFFT and 0.156 for interpolation cases. The standard deviation to mean ratio of the log spectrum is 0.33 for NUFFT and 0.36 for interpolation method in the example shown.

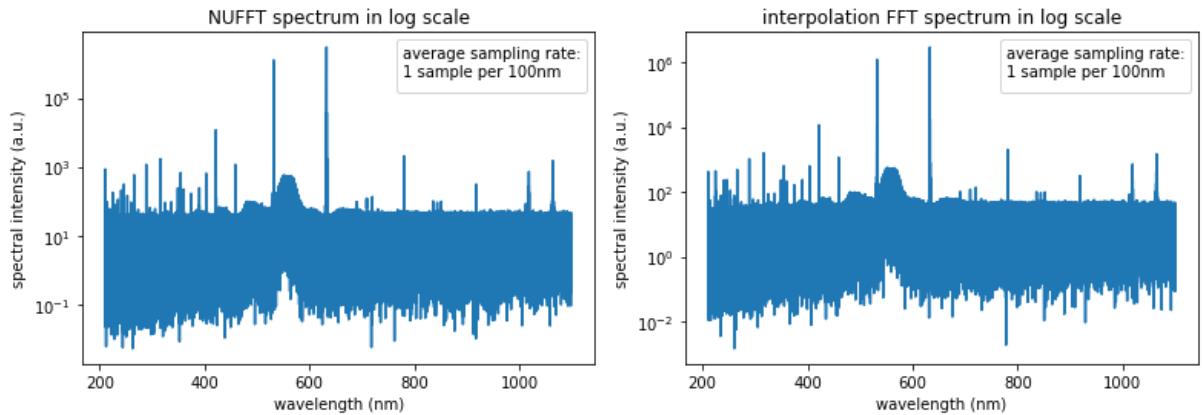

*Figure 8 The log scale spectrum of 532nm laser for NUFFT (left) and interpolation (right) methods. The method for interpolation is Cubic Spline.*

**4.1.3 Spectral amplitude**:

In section 4.1.1 it was mentioned that the spectral amplitude produced by the two methods can be significantly different. To investigate whether the amplitude difference depends on source type / spectral method, we looked at the spectral amplitude differences in more detail by investigating whether the difference is related to the sampling rate. We found that this is the case with the 532nm laser. For the 532nm laser, the NUFFT amplitude is about 0.5% higher than interpolation method when the sampling rate is about 1 per 105nm. This percentage does not change from run to run. When the sampling rate is reduced by half by omitting 1 point for every 2 sample points to about 1 per 210nm (corresponding Nyquist rate for 532nm is 1 per 266nm), the spectral amplitude of the NUFFT method is 20% higher (Figure 9) than the Interpolation method. Furthermore, the obtained NUFFT spectral amplitude is proportional to the sampling rate and thus the 20% difference is due to the interpolation method not scaling linearly with sampling rate.

Interestingly, this behaviour is not observed in the 960nm SLED case. We found that while the NUFFT amplitude is still proportional to the sampling rate, the amplitude difference with interpolation methods is inconsistent for different sets of acquired samples. Sometimes the NUFFT amplitude is higher than interpolation method while sometimes it is lower with no apparent trend. And for the same scan, reducing the sampling rate by omitting points will not change this amplitude difference. But although the amplitude difference varies with different samples, we found that the difference is always the same shape as the spectrum shape regardless of the sign. To illustrate this, we subtract one example of the NUFFT spectrum of the 960nm SLED source with the interpolation FFT spectrum (Figure 10). Thus, this proportionality ensures that the obtained spectral profiles from NUFFT and interpolation will be very similar even though the amplitudes differ slightly. The difference between cubic spline interpolation and linear interpolation is also shown for comparison showing much less difference and so the method of interpolation is not a factor.

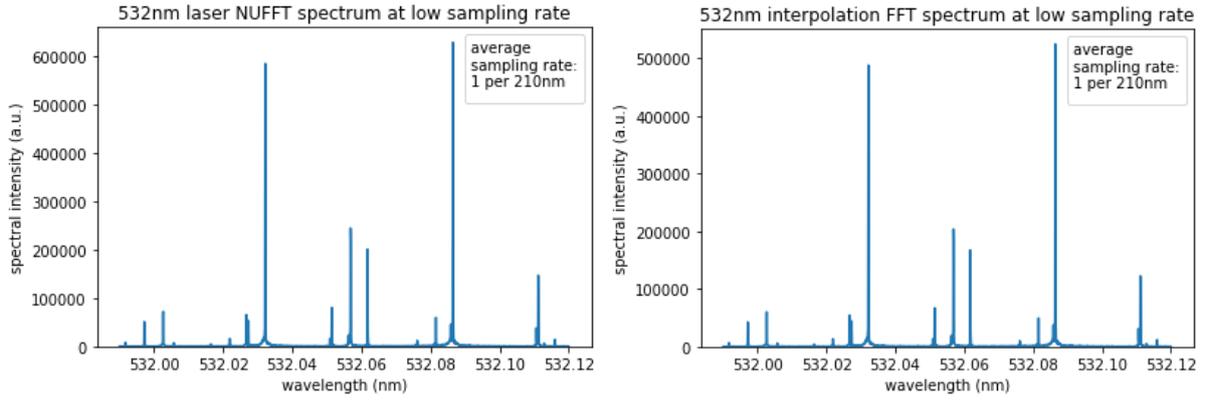

*Figure 9 The spectral amplitude difference of NUFFT (left) and interpolation FFT (right) for the 532nm laser*

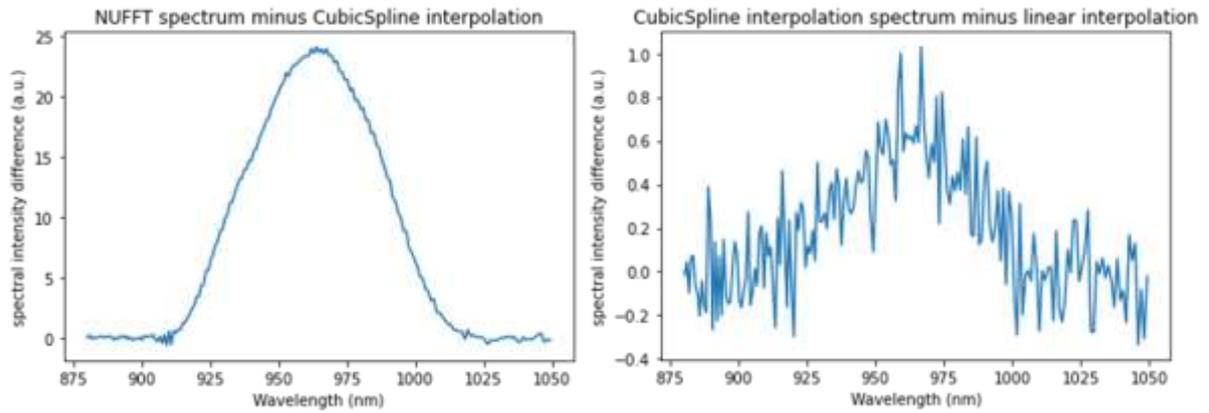

*Figure 10 The magnitude difference between the NUFFT spectrum and interpolation spectrum (left). The difference between the Cubic Spline interpolation and linear interpolation (right). The average sampling interval is about 20nm.*

Thus, we have shown that, when normalised to number of samples, in general the NUFFT amplitude is less affected by sampling rate and is larger in the 532nm laser case. The difference in behaviour of the spectral amplitude between the laser and the SLED may be due to the fact that laser spectrum is discrete while the SLED spectrum is broad and continuous.

**4.2 Under-sampling and aliasing behaviour:**

The periodic nature of the regular discrete Fourier transform equation means that the calculated spectrum has a maximum limit in frequency and any frequency higher than the limit will fold back into the spectrum thereby distorting the calculated spectrum, as can be seen from the equation. This problem is called aliasing.

$$f(n) = \sum_k F(k) e^{i2\pi \frac{nk}{N}} = \sum_k F(k) e^{i2\pi \frac{n(k+N)}{N}}$$

where $k+N$ will be the false frequency appearing in the spectrum because $k+N$ and so on is fundamentally undiscernible from $k$ in the equation.

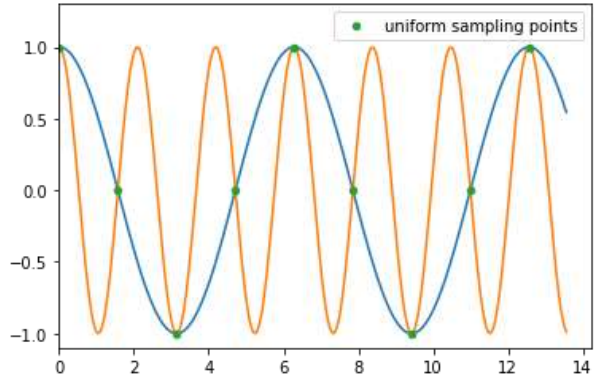

*Figure 11 An illustration of aliasing in a uniformly sampled situation. Two signals with different frequencies give same signal values at the uniform sampling points.*

By contrast, the NUFFT is inherently non-periodic since usually the sample positions are randomly distributed and so should have no maximum frequency limits and be immune to aliasing if the sample position distribution is random enough.

To illustrate this, we used a 532nm laser interferogram and reduced the sampling rate by dropping two out of three sampling points (sampling rate reduces to 1 point per 316nm from 1 point per 105nm). For the interpolation FFT spectrum the 532nm signal is folded into the spectrum and appears as a false 775nm signal while in the NUFFT case there is only an elevated noise level around the expected aliasing region due to the quasi-periodic nature of the sampling (see Figure 12). In addition, the NUFFT also has no maximum frequency limit and can still calculate the 532nm spectrum in the under-sampled case (see Figure 13 for both original and under-sampled cases).

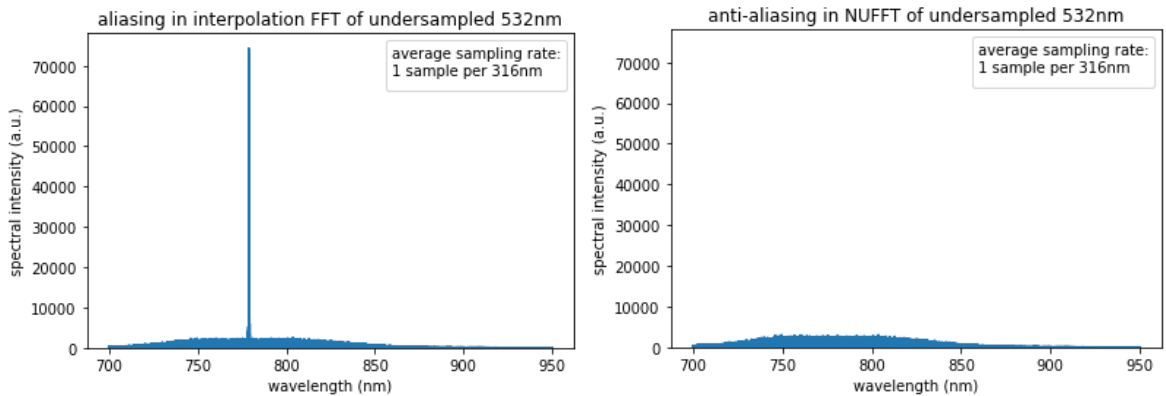

*Figure 12 Illustration of aliasing in interpolation FFT (left) and compared to NUFFT (right)*

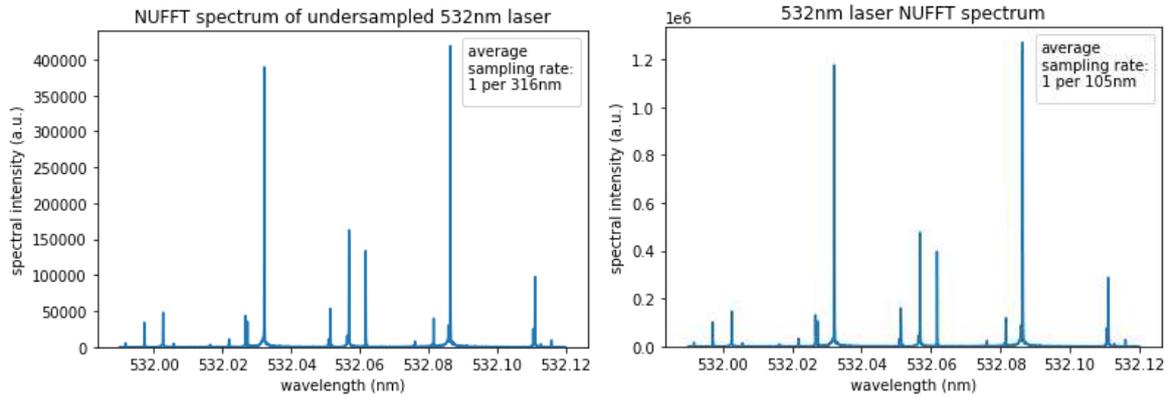

*Figure 13 Under-sampled 532nm laser spectrum (left) compared to the normally sampled one (right).*

To illustrate the effect of severe under sampling in NUFFT, we under sampled the 532nm laser by omitting 9 out of every 10 sample points from the master interferogram and calculate the spectrum by NUFFT. We also plot the spectrum in log scale to compare the noise levels (see Figure 14). It shows that the severely under-sampled interferogram can reproduce the laser spectrum as well as the original one, however its noise level is significantly higher. It's worth mentioning here that the spectral resolution is unchanged because the frequency resolution of Fourier transform spectrometers depends on the total scan length rather than the number of sampling points.

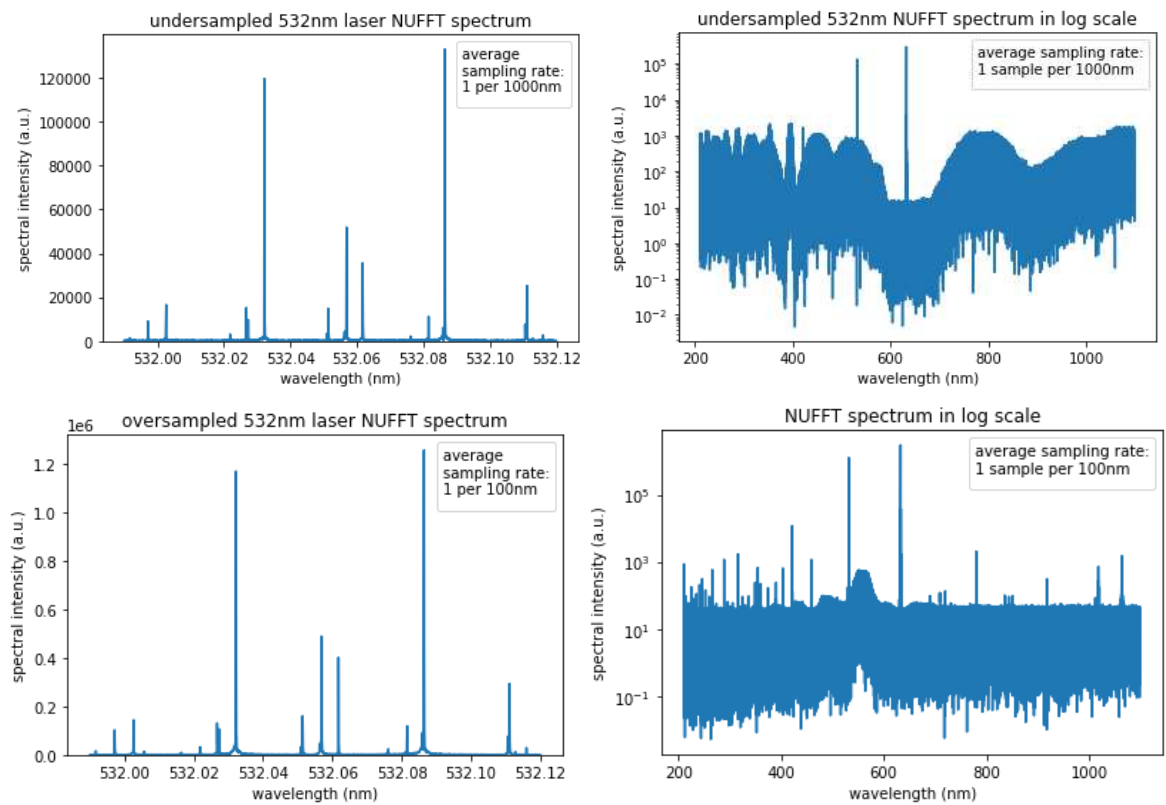

*Figure 14 An under-sampled 532nm laser spectrum plotted on linear (top left) and log scales (top right) and the corresponding oversampled cases (bottom).*

**4.3 Non-random electrical noise:**

The photodetector and the data acquisition device will contribute some electrical noise to the measured interferogram. This noise can be directly observed by measuring the signal in the absence of any light. An example of the spectrum of one such scan is shown on Figure 15 and shows the presence of some particular frequencies. In the conventional uniformly sampled interferogram cases these frequencies can contribute to the overall signal and reduce the accuracy of the final spectrum.

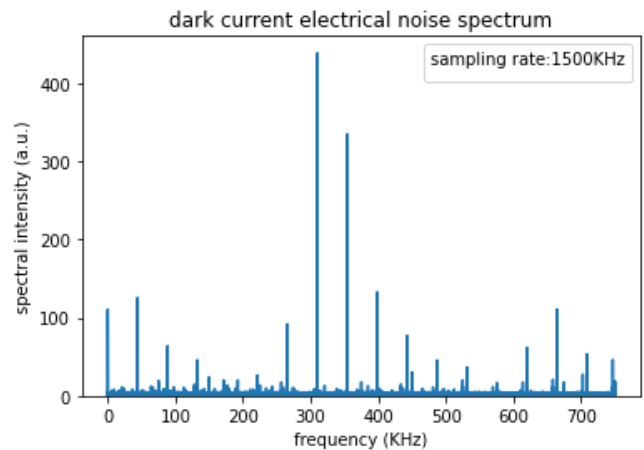

Figure 15 An example of dark current noise spectrum. Sampling rate is set at 1500KHz.

However, non-uniform sampling is inherently advantageous over uniform sampling in mitigating this kind of non-random electric noise because the electric noise exists in the time domain while the interferometric signal exists in the optical path difference domain. This property of nonuniform sampling spectrometer was tested with both NUFFT and interpolation method by running the experiment with the reference helium neon laser on and other sources off (Figure 16). The intensity of helium neon laser is reduced to make the electrical noise more prominent. It can be seen from the figure that in the time domain spectrum of the HeNe interferogram there is one broad HeNe peak plus some sharp spikes due to electrical noise. However, in the optical path domain spectra calculated by either NUFFT or interpolation FFT method, the HeNe peak sharpens while the electrical noise spikes disappear (plotted on log scale to view noise level). Both the NUFFT method and interpolation FFT method performs similarly in reducing these electrical noise periodic components because of the non-uniform sampling.

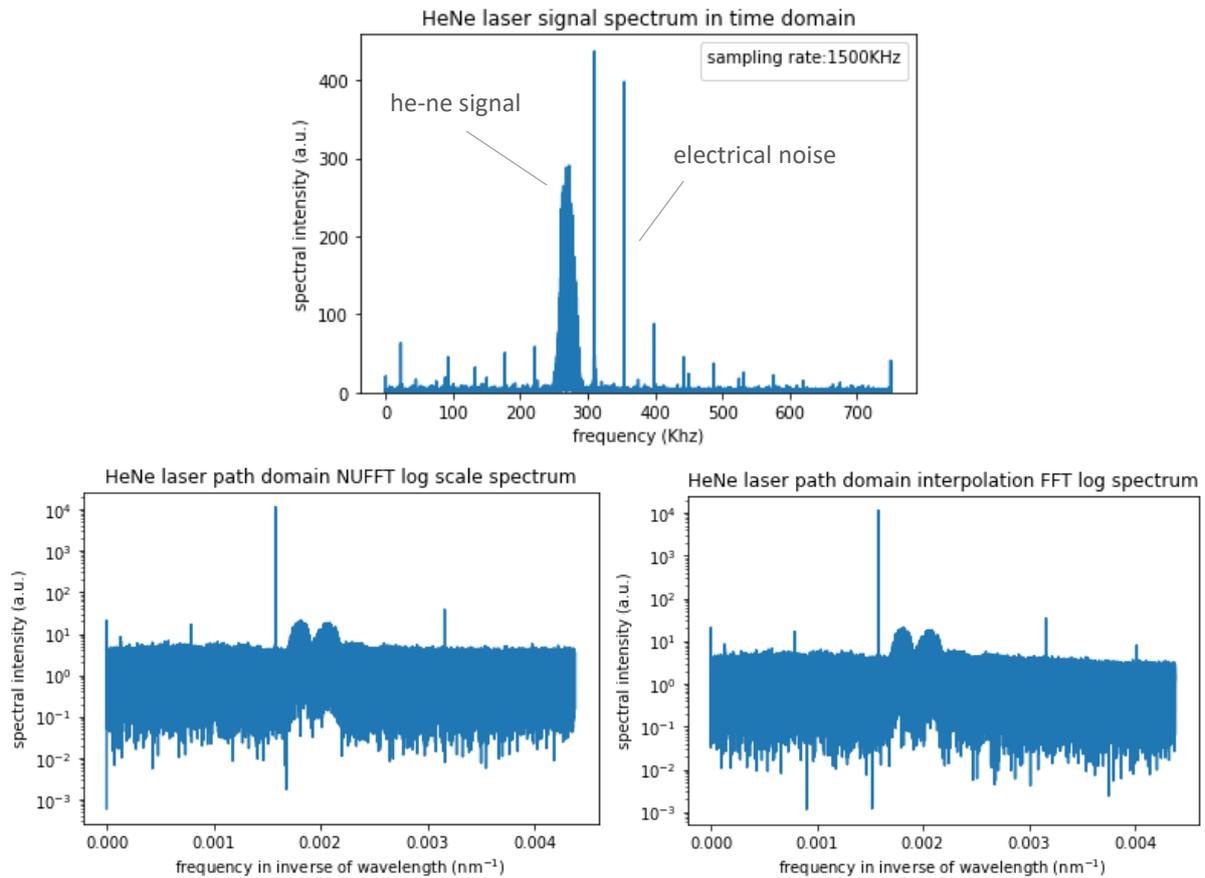

*Figure 16 Time domain spectrum (top) of a Helium neon laser interferogram and optical path difference domain spectrum from NUFFT method (bottom left) and interpolation FFT method (bottom right).*

**4.4 Computation performance comparison in practice:**

It is usually impractical to develop NUFFT or FFT software codes because those are relatively complex algorithms that would take a huge amount of time to create. Thus, the availability of corresponding software packages is very important in a practical sense. In this section we compare the software availability and computing performance of both methods in our implementation.

The availability of software for NUFFT is a relatively weak point compared to the interpolation method. Both FFT and interpolation are very mature mathematical techniques that have standard packages in many programming languages. By contrast NUFFT is a newer and less developed technique and for many software programs there are no standard software libraries for it. As a result, many existing NUFFT implementations are only made for specific applications. This makes those libraries lack usability for users from other fields. In fact, many of the NUFFT packages we have attempted to try for this study seem to be poorly maintained and lack adequate supporting documents while the NUFFT library used by this study only came to existence in 2017.

Theoretically both NUFFT and FFT speed scale up similarly with the logarithm of the number of spectral points [16,17]. Therefore, there should be no fundamental speed advantage or disadvantage by either NUFFT or interpolation methods. To investigate this issue, we carried out a rough comparison of the actual performance of our implementation of NUFFT method (FINUFFT Python package [16,17]) with our interpolation FFT method (Python SciPy interpolate and FFT packages [18]) from both a computation time and memory consumption perspective.

For CPU computing time the NUFFT method is generally as fast (and sometimes faster) than the cubic spline interpolation /FFT method. The difference depends on sample size because the speed does not scale linearly with the number of sample points. For example, for a sample size of 40 million sample points, the NUFFT is about 32 seconds while the interpolation method is around 48 seconds. However, for other sample sizes the difference can be much less. In addition, the NUFFT method consumes less memory than the interpolation / FFT method i.e. the linear interpolation / FFT method requires 50% more memory than the NUFFT and the cubic interpolation / FFT method requires 3 times as much memory.

Also, it is important to note another very significant practical advantage of the FINUFFT package in that it allows users to select only a subset of spectral points to calculate which significantly reduces computing time while standard FFT packages always compute the whole spectrum.

Thus, overall NUFFT is more efficient in our Python implementation, although the time / resource saving may be small compared to other processing procedures such as graph plotting etc. While it is entirely possible that this advantage of NUFFT is only due to better optimization by the FINUFFT package, this at least demonstrates that NUFFT has practical advantages.

## 5. Discussion

We have used experiments to compare the results of both NUFFT and interpolation FFT methods in spectral shape, spectral amplitude, spectral noise levels, aliasing and under sampling behaviour and computer performance. Next, in this section we would use simulation and theoretical derivation to better understand the findings from the experiment results. We simulate an interferogram from an ideal monochromatic source to compare the difference between both methods in spectral amplitude and noises. We also tested a non-standard form of nonuniform Fourier transform equation devised by us and shows better noise level than the standard nonuniform Fourier transform.

**5.1 Spectral noise:**

Usually an experiment contains many kinds of noises. However, we are only interested in the noise differences between the two methods. Thus, we will use simulation to better compare the spectral noises generated by the two methods. According to the equation of either nonuniform Fourier transform and uniform transform, a signal of a single frequency can introduce values to other parts of the calculated spectrum. This could be considered a type of spectral noise/error. In addition, for the interpolation FFT method the interpolation could also introduce some errors to the interpolated interferogram which would result in additional noises.

We construct the interferogram of an ideal non-uniformly sampled monochromatic 632.8nm source (essentially a cosine wave) using the sampling position information from one of the experimental measurements. We then calculate the spectrum from this simulated interferogram using both NUFFT and interpolation methods and plot the result on a log scale. Figure 17 shows the result of cubic spline interpolation, linear interpolation, NUFFT and an under-sampled NUFFT result. The average sampling rate was set to about twice the Nyquist rate so that the signal's signature spectral peak appears in the centre of the calculated spectrum. It shows that the noise level of NUFFT is comparable to interpolation, but their noise shape is significantly different, with the NUFFT shape being much more irregular. We also test the under-sampling feature of NUFFT and show that under-sampling will result in significantly higher noise levels. In addition, the figure also indicates the values of the peak of the spectrum. We have chosen the amplitude of the simulated signal such that the theoretical spectral amplitude should equal to 1. It can be seen that linear interpolation has a much lower spectral peak value than other methods while the NUFFT value agrees perfectly with the predicted one. We will discuss the theoretical spectral amplitude further in the next section.

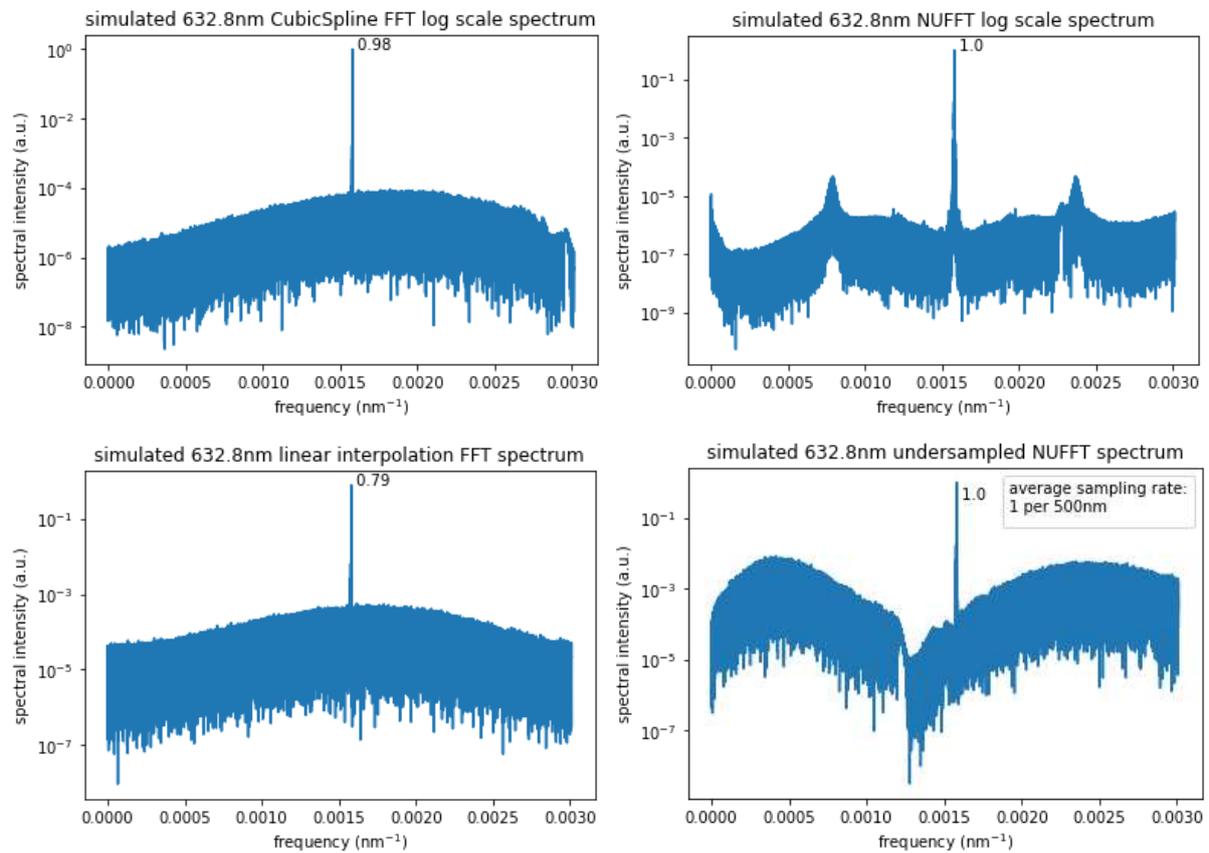

Figure 17 Comparison of simulated spectral noise by cubic spline interpolation (top left), NUFFT (top right), linear interpolation (bottom left), and under-sampled NUFFT(bottom right). The average sampling rate is about 1 per 170nm except the last graph which is under sampled at around 1 per 500nm.

**5.2 Spectral amplitude:**

In section 4.1.3 it was demonstrated that the NUFFT spectral amplitude is proportional to the average sampling rate but the interpolation FFT spectral amplitude is not linearly related to sampling rate and is also dependent on the type of optical source.

To better understand this phenomenon, we first try to determine the theoretical spectral amplitudes for both non-uniform and uniform discrete Fourier transform equation. According to the equation of both non-uniform and uniform discrete Fourier transform, the calculated spectral amplitude for an ideal monochromatic signal without normalization is equal to the signal amplitude times the number of sample points if the signal frequency happens to fall on one of the spectral points calculated i.e.

$$F(v_k) = \sum_{n=1}^{N}(Ae^{i2\pi v x_n})e^{-i2\pi v_k x_n} = NA \ \ for \ v_k = v$$

where $Ae^{i2\pi v x}$ is the monochromatic source and $A, v$ are the signal amplitude and frequency respectively. When spectral noise is considered, the above equation may not be true. However, the level of noise in the experiment is not sufficient to account for the differences between NUFFT and interpolation methods.

Combined with the fact that the NUFFT spectral amplitude is proportional to sampling rate in the experimental result. We can postulate that the amplitude difference between the two methods is due to the interpolated signal being not the same as the actual signal. Suppose that on the contrary interpolation can perfectly reproduce the actual signal value, then we should expect no amplitude difference in the calculated results. For the 532nm laser source the experiment result shows that the interpolated signal amplitude decreases with decrease of sampling rate. For the 960nm SLED source the result shows that the interpolated signal amplitude can be larger or smaller than the actual amplitude with no apparent trend. We postulate that this is due to that the interferogram of a broadband source is very sharp, its amplitude is strongest when the optical path difference equals 0 and quickly loses strength as the optical path difference increases. Thus, the position of interpolation points and sample points may cause the amplitude of the interpolated interferogram to be either overestimated or underestimated.

**5.3 A non-standard variant of NUFFT:**

Non-uniform discrete Fourier transform was devised to approximate the continuous Fourier transform. As far as we know, there are no fundamental reasons why the equation for the non-uniform Fourier transform has to be the way as it is defined and perhaps there can be other definitions that would be more suitable for Fourier transform spectrometry applications. To explore this point, we devised and tested a nonstandard variant of the non-uniform Fourier transform equation after noticing a discrepancy that the regular discrete Fourier transform would converge to the formula of continuous Fourier transform when the sampling rate increases but the standard non-uniform Fourier transform would not to see whether addressing this discrepancy by giving each sample points a weight based on the local sampling interval can give better spectral results. The definition is given below:

$$F(v_n) \equiv \sum_{k=1}^{N} f(x_k) \frac{(x_{k+1} - x_{k-1})}{2} e^{-i2\pi v_n x_k}$$

This modified summation formula of non-uniform Fourier transform will converge to the integration formula of the continuous Fourier transform when the average sampling rate is sufficiently high:

$$\sum_{k=1}^{N} f(x_k) \frac{(x_{k+1} - x_{k-1})}{2} e^{-i2\pi v_n x_k} \xrightarrow{\Delta x \to 0} \int_{x_0}^{x_N} f(x) e^{-i2\pi v_n x} dx$$

A positive aspect of this modification is that almost no speed penalty will be incurred by its inclusion and the existing NUFFT package can still be utilised. To clearly illustrate the effect of this modification, a similarly constructed interferogram is utilised again here (shown on Figure 18 where the average sampling rate was set at about 1 per 125nm and the frequency is shown in inverse of wavelength). The corresponding maximum frequency limit (Nyquist frequency) in the uniformly sampling discrete Fourier transform would have been 250nm or 0.004nm$^{-1}$. But as outlined in section 4.2 the NUFFT has no maximum frequency limit. It can be seen from Figure 18 that this nonstandard variant of NUFFT has a significantly lower spectral noise level for frequencies lower than the Nyquist frequency (<0.004nm$^{-1}$). However, it shows no improvement in the under sampled region (>0.004nm$^{-1}$). This demonstrates again that even though there are no theoretical maximum frequency limits to the NUFFT there are still behavioural differences between under sampled and sufficiently sampled region. Testing with the experimental data shows that the difference with standard NUFFT is very minimal with marginal improvements.

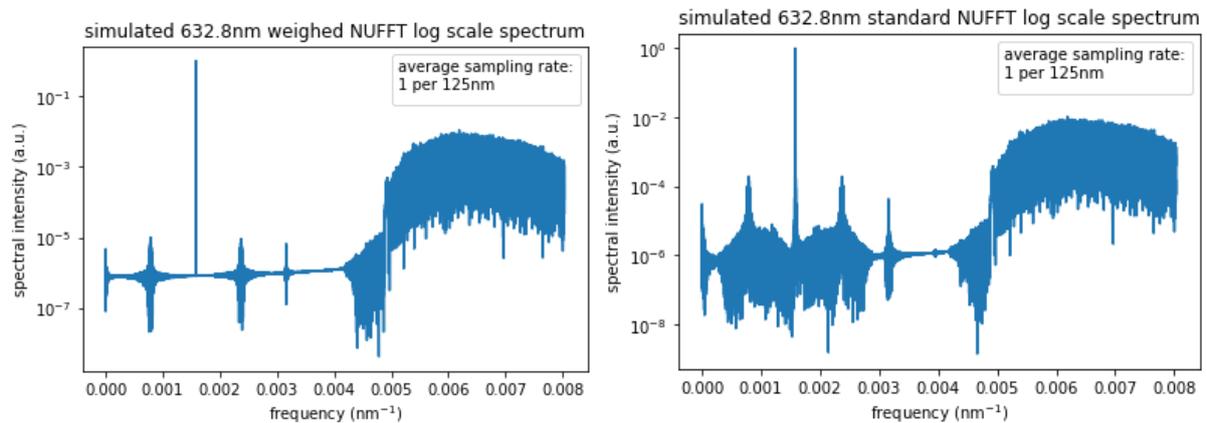

Figure 18 Comparison of spectrum of an ideal 632.8nm cosine wave signal by weighted NUFFT (left) to that of standard NUFFT (right).

# 6. Conclusion

A detailed experimental study has been carried out to show that the NUFFT method is as good (and sometime better) than the resampling by interpolation / FFT method for a non-uniformly sampled Fourier transform spectrometer in terms of spectral shape and spectral noise levels. The NUFFT method is also better in reproducing spectral amplitudes and has advantages in under-sampled situations where aliasing may be an issue. It was also shown that non-uniform sampling has fundamental benefits over uniform sampling due to its non-periodic nature with under sampling, anti-aliasing and electric noise mitigation properties. The NUFFT method also consumes less computer memory and can be faster than interpolation FFT in certain cases. A nonstandard form of NUFFT was also investigated and showed marginal improvement over the original standard form of non-uniform Fourier transform. As a result, it can be concluded that NUFFT is an equal or superior alternative to traditional interpolation / FFT method for a non-uniformly sampled Fourier transform spectrometer.

# Acknowledgement

This research is supported by South East Technological University (Waterford Institute of Technology), Ireland.

# Reference

1. Davis, Sumner P., Mark C. Abrams, and James W. Brault. Fourier transform spectrometry. Elsevier, 2001.
2. de Oliveira, Nelson, et al. "High-resolution broad-bandwidth Fourier-transform absorption spectroscopy in the VUV range down to 40 nm." Nature photonics 5.3 (2011): 149-153. https://doi.org/10.1038/nphoton.2010.314
3. Meng, Yijian, et al. "Interferometric time delay correction for Fourier transform spectroscopy in the extreme ultraviolet." Journal of Modern Optics 63.17 (2016): 1661-1667. DOI:10.1080/09500340.2016.1165872
4. Marvasti, Farokh, ed. Nonuniform sampling: theory and practice. Springer Science & Business Media, 2012.

5. Bondesson, David, et al. "Nonuniform Fourier‐decomposition MRI for ventilation‐and perfusion‐weighted imaging of the lung." Magnetic Resonance in Medicine 82.4 (2019): 1312-1321. https://doi.org/10.1002/mrm.27803
6. Kruizinga, Pieter, et al. "Plane-wave ultrasound beamforming using a nonuniform fast Fourier transform." IEEE transactions on ultrasonics, ferroelectrics, and frequency control 59.12 (2012): 2684-2691. doi: 10.1109/TUFFC.2012.2509
7. Salehi-Barzegar, Alireza, et al. "A fast diffraction tomography algorithm for 3-D through-the-wall radar imaging using nonuniform fast Fourier transform." IEEE Geoscience and Remote Sensing Letters 19 (2020): 1-5. doi: 10.1109/LGRS.2020.3021793
8. Liu, Qing Huo, et al. "Applications of nonuniform fast transform algorithms in numerical solutions of differential and integral equations." IEEE Transactions on geoscience and remote sensing 38.4 (2000): 1551-1560. doi: 10.1109/36.851955
9. Bagchi, Sonali, and Sanjit K. Mitra. "The nonuniform discrete Fourier transform and its applications in filter design. I. 1-D." IEEE Transactions on Circuits and Systems II: Analog and Digital Signal Processing 43.6 (1996): 422-433. doi: 10.1109/82.502315.
10. Yang, Zhengfan, and Pawel A. Penczek. "Cryo-EM image alignment based on nonuniform fast Fourier transform." Ultramicroscopy 108.9 (2008): 959-969. https://doi.org/10.1016/j.ultramic.2008.03.006
11. Chan, Kenny KH, and Shuo Tang. "High-speed spectral domain optical coherence tomography using non-uniform fast Fourier transform." Biomedical optics express 1.5 (2010): 1309-1319.
12. Shimobaba, Tomoyoshi, et al. "Nonuniform sampled scalar diffraction calculation using nonuniform fast Fourier transform." Optics letters 38.23 (2013): 5130-5133.
13. Guo, Renhui, et al. "Optical homogeneity measurement of parallel plates by wavelength-tuning interferometry using nonuniform fast Fourier transform." Optics Express 27.9 (2019): 13072-13082.
14. Schardt, Michael, et al. "Static Fourier transform infrared spectrometer." Optics Express 24.7 (2016): 7767-7776.
15. Naylor, David A., et al. "Data processing pipeline for a time-sampled imaging Fourier transform spectrometer." Imaging Spectrometry X. Vol. 5546. SPIE, 2004. https://doi.org/10.1117/12.560096
16. Barnett, Alexander H., Jeremy Magland, and Ludvig af Klinteberg. "A parallel nonuniform fast Fourier transform library based on an "Exponential of semicircle" kernel." SIAM Journal on Scientific Computing 41.5 (2019): C479-C504.
17. Barnett, Alex H. "Aliasing error of the $exp(\beta\sqrt{1-z^2})$ kernel in the nonuniform fast Fourier transform." Applied and Computational Harmonic Analysis 51 (2021): 1-16.
18. Virtanen, Pauli, et al. "SciPy 1.0: fundamental algorithms for scientific computing in Python." Nature methods 17.3 (2020): 261-272.
19. Kauppinen, J., T. Kärkkäinen, and E. Kyrö. "Correcting errors in the optical path difference in Fourier spectroscopy: a new accurate method." Applied Optics 17.10 (1978): 1587-1594.
20. Ahro, Mikko, Jyrki Kauppinen, and Ilkka Salomaa. "Detection and correction of instrumental line-shape distortions in Fourier spectroscopy." Applied Optics 39.33 (2000): 6230-6237.
21. Brault, James W. "New approach to high-precision Fourier transform spectrometer design." Applied Optics 35.16 (1996): 2891-2896.